\documentclass[a4paper]{article}

\usepackage[english]{babel}
\usepackage[utf8]{inputenc}
\usepackage{amsmath}
\usepackage{graphicx}
\usepackage[colorinlistoftodos]{todonotes}
\usepackage{authblk}
\usepackage{lineno}

\title{Deep Neural Networks Reveal a Gradient in the Complexity of Neural Representations across the Brain’s Ventral Visual Pathway}

\author[1]{Umut G\"{u}\c{c}l\"{u}%
  \thanks{Correspondence to: \texttt{u.guclu@donders.ru.nl}}}
\author[1]{Marcel A. J. van Gerven}
\affil[1]{Radboud University Nijmegen, Donders Institute for Brain, Cognition and Behaviour, Nijmegen, the Netherlands}

\date{\today}

\begin{document}
%\linenumbers
\maketitle

\begin{abstract}
Converging evidence suggests that the mammalian ventral visual pathway encodes increasingly complex stimulus features in downstream areas. Using deep convolutional neural networks, we can now quantitatively demonstrate that there is indeed an explicit gradient for feature complexity in the ventral pathway of the human brain. Our approach also allows stimulus features of increasing complexity to be mapped across the human brain, providing an automated approach to probing how representations are mapped across the cortical sheet. Finally, it is shown that deep convolutional neural networks allow decoding of representations in the human brain at a previously unattainable degree of accuracy, providing a more sensitive window into the human brain.
\end{abstract}

\section{Introduction}
\label{sec:introduction}

Human beings are extremely adept at recognizing complex objects based on elementary visual sensations. Object recognition appears to be solved in the mammalian brain via a cascade of neural computations along the visual ventral stream that represents increasingly complex stimulus features, which derive from the retinal input \cite{Kobatake1994}. That is, neurons in early visual areas have small receptive fields and respond to simple features such as edge orientation \cite{Hubel1962}, whereas neurons further along the ventral pathway have larger receptive fields, are more invariant to transformations and can be selective for complex shapes \cite{Hung2005}. 

Despite converging evidence concerning the steady progression in feature complexity along the ventral stream, this progression has never been properly quantified across multiple regions in the human ventral stream. Furthermore, while the receptive fields in early visual area V1 have been properly characterized in terms of preferred orientation, location and spatial frequency \cite{Jones1987}, exactly what stimulus features are represented in downstream areas is more heavily debated \cite{Grill2009}.  

In order to isolate how stimulus features at different representational complexities are represented across the cortical sheet, we made use of a deep convolutional neural network (CNN). Deep CNNs consist of multiple layers where deeper layers can be shown to respond to increasingly complex stimulus features and provide state-of-the-art object recognition performance in computer vision \cite{Krizhevsky2012}. We used the representations that emerge after training a deep CNN in order to predict blood-oxygen-level dependent (BOLD) hemodynamic responses to complex naturalistic stimuli in progressively downstream areas of the ventral stream, moving from striate area V1 along extrastriate areas V2 and V4, all the way up to area LOC in posterior inferior temporal (IT) cortex. 

We used individual layers of the neural network to predict single voxel responses to natural images. This allowed us to isolate different voxel groups, whose responses are best predicted by a particular layer in the neural network. Using this approach, we can determine how layer depth correlates with the position of voxels in the visual hierarchy. Furthermore, by testing to what extent individual features in the neural network can predict voxel responses, we can map how individual low-, mid- and high-level stimulus features are represented across the ventral stream. This provides a unique and fully automated approach to determine how stimulus features of increasing complexity are represented across the visual stream. Finally, we show that the predictions of neural responses afforded by our framework give rise to state-of-the-art decoding performance, allowing identification of perceived stimuli from observed BOLD responses.

\section{Framework}
\label{sec:framework}

\begin{figure}[t]
\centering
\includegraphics[width=1\textwidth]{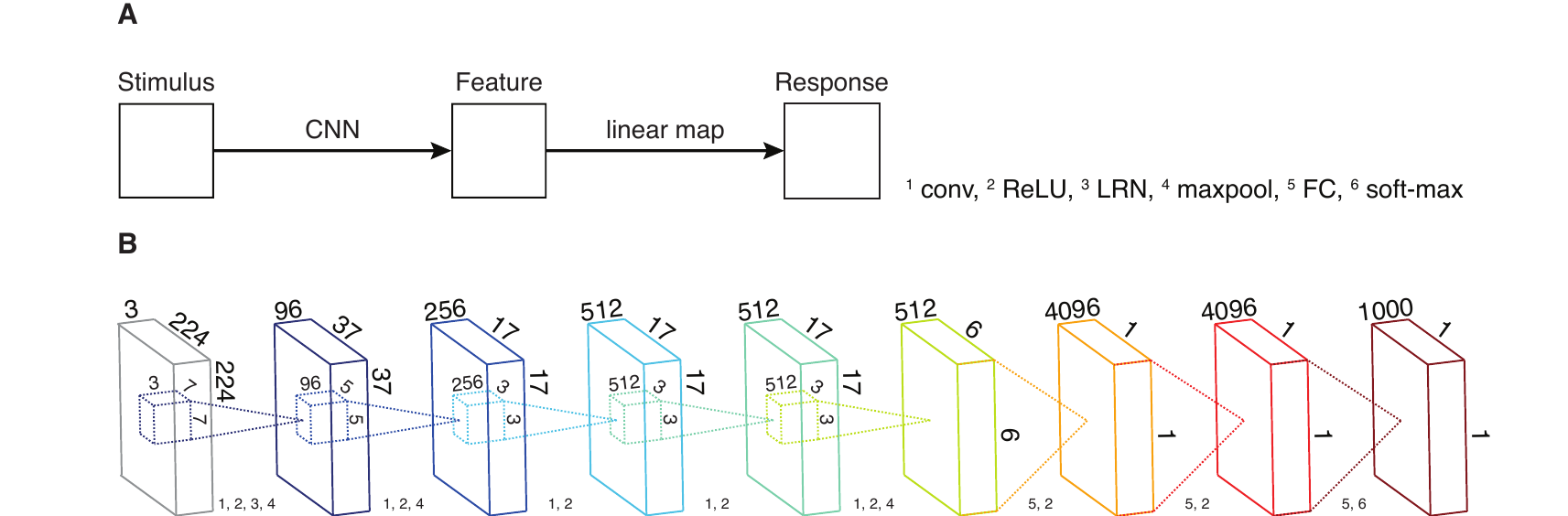}
\caption{\label{fig:figure_1}Framework. \textbf{(A)} Schematic of the encoding model. The encoding model transforms a visual stimulus to a voxel response in two stages. First, the deep convolutional neural network (CNN) transforms the visual stimulus to a layer of feature representations. Then, a linear map transforms the layer of feature representations to a voxel response. \textbf{(B)} Schematic of the deep CNN.  The deep CNN transforms a visual stimulus to different layers of feature representations. It has five convolutional and three fully connected layers of artificial neurons. Each artificial neuron in the convolutional layers repeatedly filters its input across overlapping subregions and forms a feature map. Each layer of artificial neurons has one or more of the following nonlinearities: rectification, local response normalization, max pooling and softmax transformation.}
\end{figure}

We use an encoding model that comprises two main components (Fig. \ref{fig:figure_1}). The first component is a nonlinear feature model that transforms a visual stimulus to different layers of feature representations. The second component is a linear response model that transforms a layer of feature representations to a voxel response.

The feature model is a deep convolutional neural network (CNN). A deep CNN is a feedforward artificial neural network that has multiple nonlinear layers of artificial neurons. The architecture of the feature model is the same as the CNN-S architecture in \cite{Chatfield2014}. It has five convolutional and three fully connected layers. Each artificial neuron in the convolutional layers corresponds to a feature detector that is independent of spatial location. It repeatedly filters its input across overlapping subregions and forms a feature map, i.e. a representation of the feature across space. Each layer of artificial neurons has one or more of the following nonlinearities: rectification, local response normalization, max pooling and softmax transformation. The feature model was trained on roughly 1.2 million natural images that are labelled with 1000 object categories by gradient descent with momentum. The Caffe framework was used to train the feature model \cite{Jia2014}. The natural images were taken from the ImageNet data set \cite{Deng2009}.

The response model is a regularized linear regression. A separate response model was trained for each voxel that takes one of the eight layers of feature representations as its input. The dimensionality of the layers one through eight was 131424, 73987, 147968, 147968, 18432, 4096, 4096, 1000, respectively. The response models were trained on 1750 stimulus-response pairs (i.e. training set) by ridge regression, and tested on 120 stimulus-response pairs (i.e. test set). The stimulus-response pairs were taken from the vim-1 data set \cite{Kay2011} that was originally published in \cite{Kay2008, Naselaris2009}. The stimulus-response pairs consist of grayscale natural images spanning 20 $\times$ 20 degrees of visual angle and stimulus-evoked peak BOLD hemodynamic responses of 25915 voxels in the occipital cortex of one subject (i.e. Subject 1). The details of the experimental procedures are presented in \cite{Kay2008}. Unless otherwise stated, all significance levels are \textit{p} $\leq$ 0.01 and Bonferroni corrected for multiple comparisons when required.

\section{Results}
\label{sec:results}

\begin{figure}[t]
\centering
\includegraphics[width=1\textwidth]{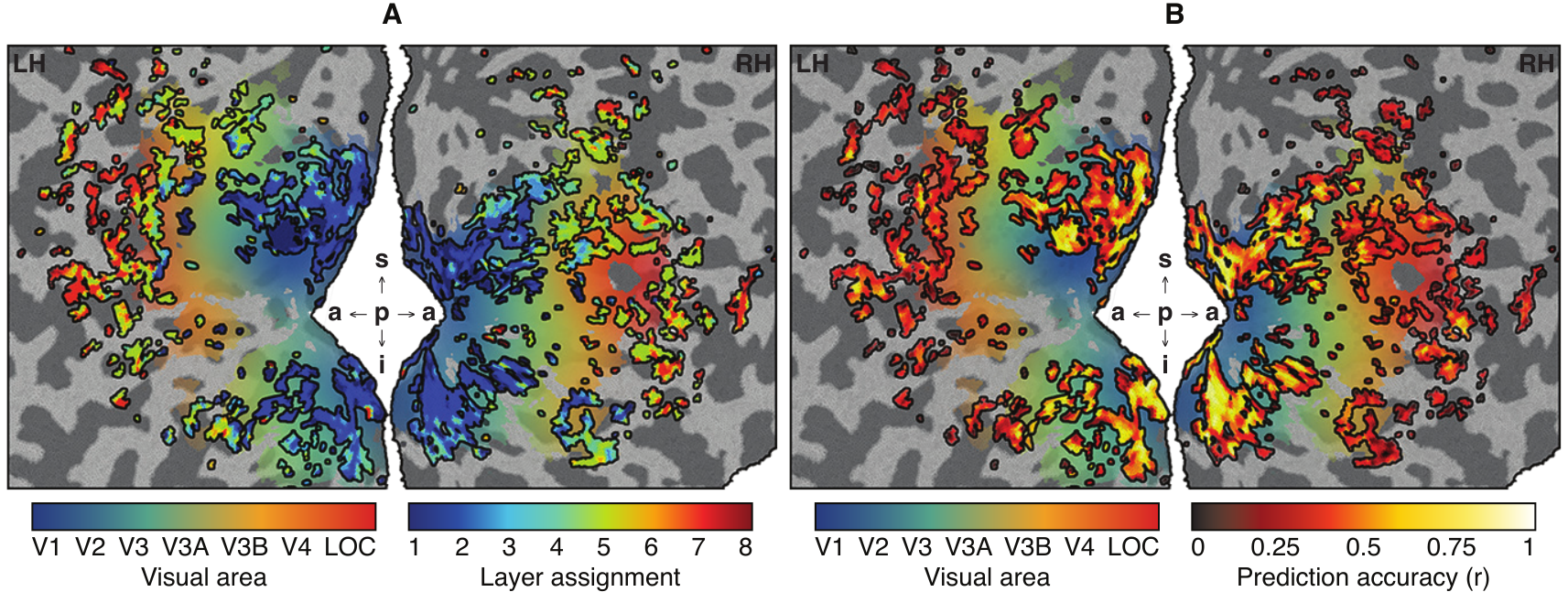}
\caption{\label{fig:figure_2}Encoding results of the significant voxels across the cortical surface. LH, RH, p, a, s and i denote left hemisphere, right hemisphere, posterior, anterior, superior and inferior, respectively. (\textbf{A}) Layer assignment of the voxels across the cortical surface. Each voxel is assigned to the layer of the deep CNN that resulted in the lowest cross-validation error on the training set. (\textbf{B}) Prediction accuracy of the voxels across the cortical surface. The prediction accuracy of a voxel is defined as the Pearson product-moment correlation coefficient (\textit{r}) between its observed and predicted responses on the test set.}
\end{figure}

We used five-fold cross-validation to assign each of the 25915 voxels to one of the eight layers of the deep CNN (Fig. \ref{fig:figure_2}A). That is, each voxel was assigned to the layer of the deep CNN that resulted in the lowest cross-validation error on the training set. Those voxels whose prediction accuracy was not significantly better than chance were discarded. The reason for nonsignificant prediction accuracy of these voxels could be either their low signal-to-noise ratio (SNR) or that none of the layers of the deep CNN reproduced their behavior. As a result, 13\% of the voxels in the occipital cortex were further analyzed.  The response models of these voxels were trained on the entire training set and evaluated on the test set. The prediction accuracy of a voxel was defined as the Pearson product-moment correlation coefficient (\textit{r}) between its observed and predicted responses on the test set. For a group of voxels, the median correlation coefficient was used to express its prediction accuracy. 

We grouped the voxels that were assigned to the same layer. While the prediction accuracy of each of the voxel groups was significantly above zero, it decreased from low- to high-layer voxel groups (Fig. \ref{fig:figure_2}B). The prediction accuracy of the voxel groups one through eight was 0.42, 0.50, 0.39, 0.29, 0.27, 0.24, 0.27 and 0.16, respectively (SE $\leq$ 0.02). The prediction accuracy was significantly correlated with the mean activity of the layers across the training set and the SNR of the voxels. This suggests that the difference in the prediction accuracy of the low- and high-layer voxel groups can be explained by the differences in the mean activity of the layers and the SNR of the voxels.

\begin{figure}[t]
\centering
\includegraphics[width=1\textwidth]{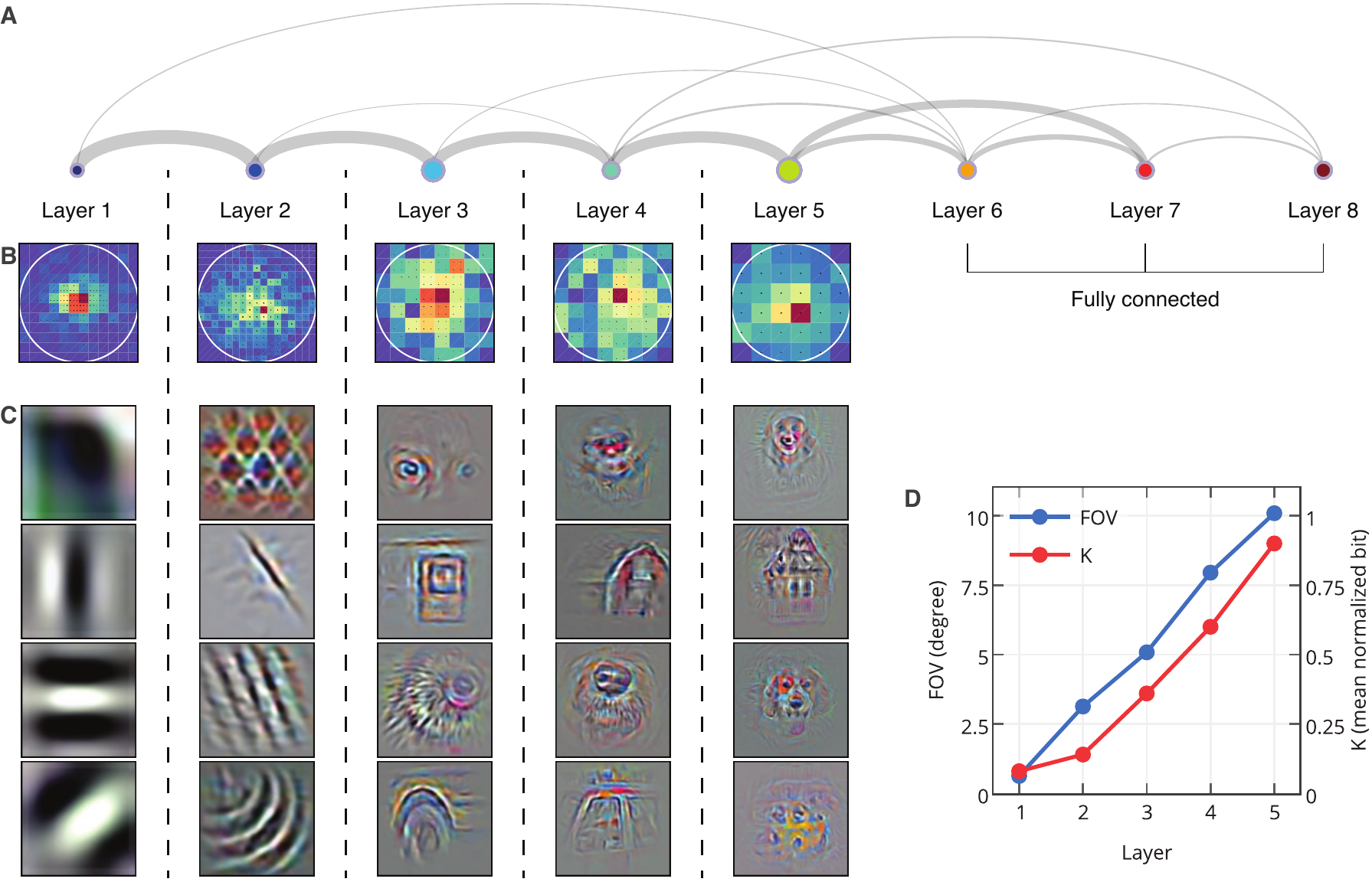}
\caption{\label{fig:figure_3}Properties of the voxel groups. (\textbf{A}) Partial correlations between the predicted responses of each pair of voxel groups, controlling for the predicted responses of the remaining voxel groups. The width of the lines are proportional to the partial correlations. (\textbf{B}) Distribution of the receptive field locations. (\textbf{C}) Examples of the internal representations. The internal representations are visualized using a deconvolutional network \cite{Zeiler2013}. (\textbf{D}) Mean field of view (FOV) and Kolmogorov complexity (\textit{K}) of the internal representations. FOV is taken to be the size of the filters. \textit{K} is taken to be the compressed file size of the internal representations.}
\end{figure}

Different voxel groups were systematically clustered around different points on the cortical surface such that an increase in the layer of the voxel groups was observed when moving from posterior to anterior points on the cortical surface. The responses of the successive voxel groups were more partially correlated than those of the non-successive voxels groups (Fig. \ref{fig:figure_3}A). The receptive fields of the voxels in each voxel group covered almost the entire field of view, with more voxels dedicated to foveal than peripheral vision (Fig. \ref{fig:figure_3}B). While there was a degree of overlap between the internal representations of the successive voxel groups, those of the low-layers resembled Gabor wavelets and textures, and those of the high-layers resembled object parts and objects (Fig. \ref{fig:figure_3}C). The mean Kolmogorov complexity (\textit{K}) of the internal representations was significantly correlated with their layer assignment (Fig. \ref{fig:figure_3}D). Taken together, these results suggest that i) each voxel group contains almost a full representation of visual space, ii) visual information travels mostly between neighboring voxel groups, and iii) moving along the voxel groups, their receptive fields increase in size, latency and complexity.

\begin{figure}[t]
\centering
\includegraphics[width=0.67\textwidth]{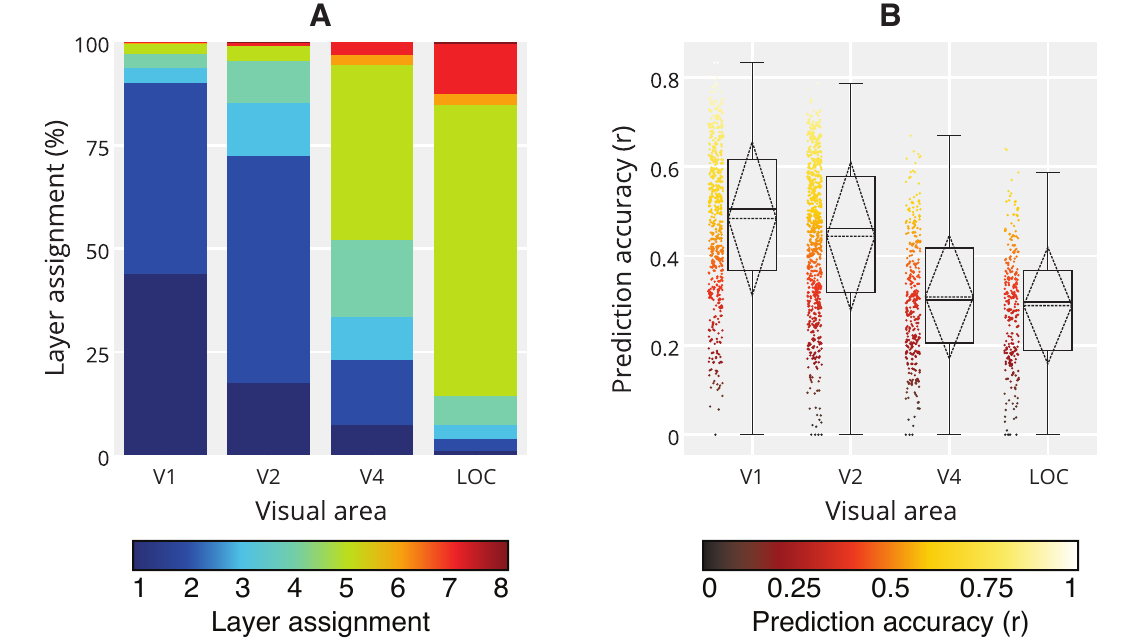}
\caption{\label{fig:figure_4}Encoding results of the significant voxels across V1, V2, V4 and LOC. (\textbf{A}) Layer assignment of the voxels across V1, V2, V4 and LOC. Each voxel is assigned to the layer of the deep CNN that resulted in the lowest cross-validation error on the training set. (\textbf{B}) Prediction accuracy of the voxels across V1, V2, V4 and LOC. The prediction accuracy of a voxel is defined as the Pearson product-moment correlation coefficient (\textit{r}) between its observed and predicted responses on the test set.}
\end{figure}

Given that these properties resemble those of the visual areas on the main afferent pathway of the ventral stream \cite{Zhaoping2014}, it is interesting to consider how these voxel groups are distributed across V1, V2, V4 and LOC. We found a systematic overlap between these voxel groups and visual areas (Fig. \ref{fig:figure_4}A). The mean layer assignment of the V1, V2, V4 and LOC voxels was 1.8, 2.3, 3.0 and 4.9, respectively (SE $\leq$ 0.1). That is, most of the low-layer voxels were located in early visual areas, whereas most of the high-layer voxels were located in downstream visual areas. Most of the fully connected voxels were located in visual areas anterior to LOC. The prediction accuracy of the V1, V2, V4 and LOC voxels was 0.51, 0.46, 0.30 and 0.30, respectively (SE $\leq$ 0.02) (Fig. \ref{fig:figure_4}B). That of the remaining voxels were 0.28 (SE $\leq$ 0.01). However, in contrast to the 30\% of the V1, V2, V4 and LOC voxels that were significant, only 8\% of the remaining voxels were significant. These results suggest that this deep CNN reproduces the behavior of the visual areas on the main afferent pathway of the ventral stream.

\begin{figure}[t]
\centering
\includegraphics[width=1\textwidth]{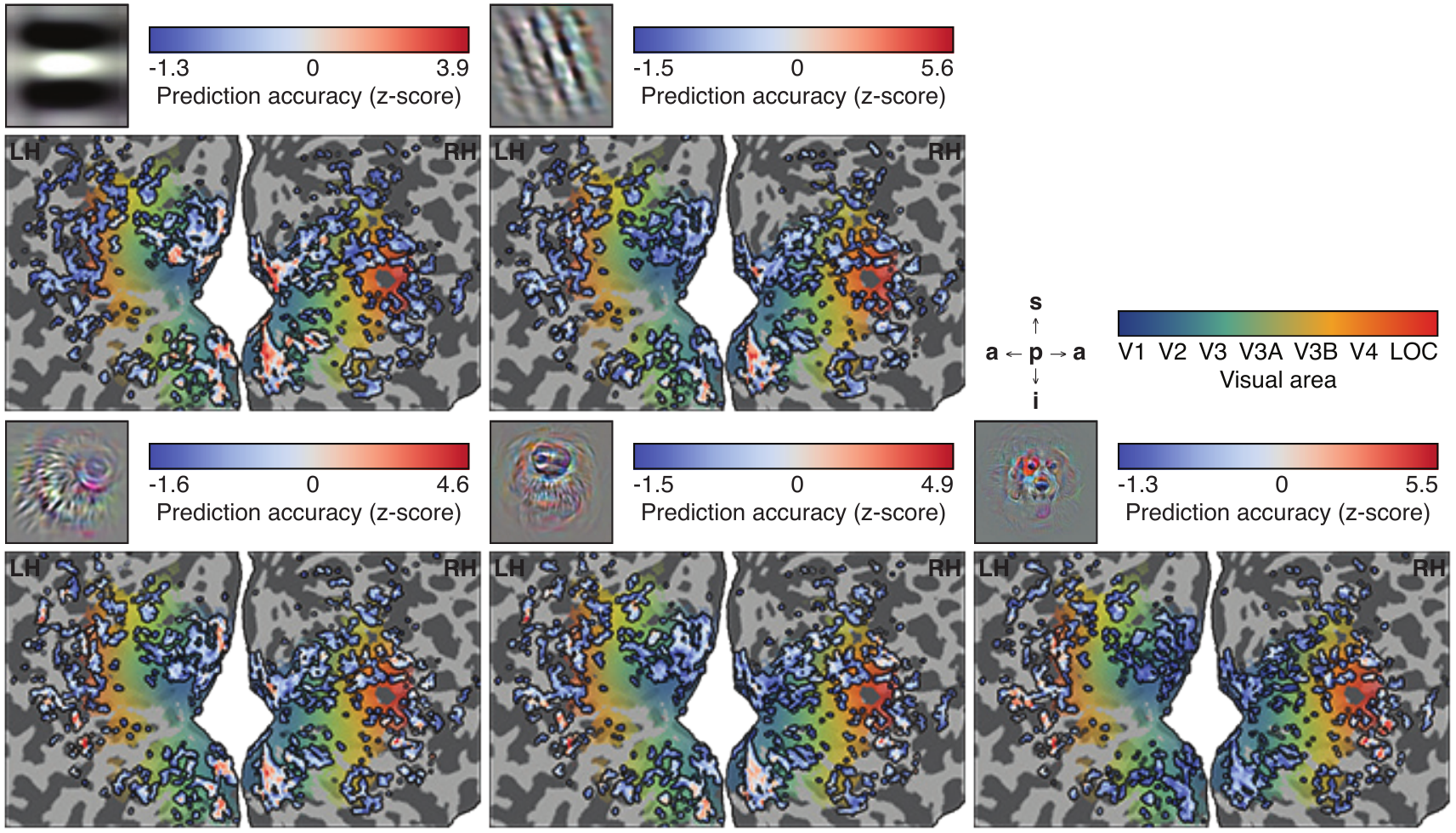}
\caption{\label{fig:figure_5}Selectivity of the significant voxels across the cortical surface to five individual feature maps. LH, RH, p, a, s and i denote left hemisphere, right hemisphere, posterior, anterior, superior and inferior, respectively.  There is a many-to-many relationship between individual feature maps and voxels that is mostly confined to single or neighboring visual areas for lowest and highest layer feature maps such as the Gabor wavelet and dog face.}
\end{figure}

To investigate how individual feature maps are represented across the cortical surface, we retrained a separate response model for each feature map - voxel combination. We quantified the selectivity of an individual voxel to an individual feature map as its corresponding prediction accuracy. We found a many-to-many relationship between individual feature maps and voxels (Fig. \ref{fig:figure_5}). That is, no individual feature map accurately predicted only one voxel, and no individual voxel was accurately predicted by only one feature map. This relationship was mostly confined to single or neighboring visual areas for highest and lowest layer feature maps. For example, a layer one feature map predicted multiple voxels exclusively in early visual areas with above average accuracy, whereas a layer five feature map predicted multiple voxels exclusively in downstream visual areas with above average accuracy.

Given the highly significant accuracy with which individual voxel responses can be predicted, it is natural to ask to what extent the deep model allows decoding of a perceived stimulus from observed multiple voxel responses alone. To answer this question, we evaluated three decoding models: a low-level (V1 + V2), a high-level (V4 + LOC) and a combined (V1 + V2 + V4 + LOC) decoding model. Given observed multiple voxel responses, the low-level decoding model correctly identified a stimulus from a set of 120 potential stimuli at 98\% accuracy, whereas the high-level decoding model correctly identified a stimulus from the same set of potential stimuli at 55\% accuracy. As the number of potential stimuli was increased from 120 to 1870, the identification performance of the low- and high-level decoding models decreased to an accuracy of 95\% and 38\%, respectively. The difference between the identification performance of the low- and high-level decoding models is not surprising since it would be more likely for two different stimuli to have ambiguously similar high-level representations than low-level representations. In fact, when we analyzed the misidentified stimuli, we found that the high-level model could most of the times identify a potential stimulus that is semantically but not structurally close to the target stimulus.

This result suggests that the combination of the low- and high-level decoding models would have a higher identification accuracy since the higher level voxels can be used to resolve the ambiguities in the feature representations of the lower level voxels and vice versa. As expected, the combined decoding model had a higher identification accuracy than either of the low- and high-level decoding models alone. It identified the correct stimulus from a set of 120 potential stimuli at 100\% accuracy. As the number of potential stimuli was increased almost 16-fold, there was no decrease in the identification accuracy. This result is a significant improvement on the earlier results in the literature where low-level features were used \cite{Kay2008, Guclu2014}, suggesting that mid- and high-level features are also important for identification.

\section{Discussion}
\label{sec:discussion}

Using a novel computational approach, we revealed a gradient in the complexity of neural representations across the brain’s ventral visual pathway. That is, it was established that downstream areas in the ventral stream code for increasingly complex stimulus features that correspond to features in increasingly deep layers of a deep convolutional neural network. This can be observed in the assignment of voxels in successive visual areas to increasingly deep layers of the neural network. In conjunction with a partial correlation analysis, which shows that information flow mainly takes place between neighboring visual areas, this result provides strong evidence for the thesis that the visual ventral stream can be seen as a hierarchical system whose downstream areas process increasingly complex features of the retinal input.

The representations that were learned by the deep neural network also allow probing of how individual stimulus features are represented across the cortical sheet. These results again revealed that low-level stimulus properties are mainly confined to early visual areas, whereas more semantically meaningful high-level stimulus properties such as object parts and objects were mostly represented in posterior inferior temporal areas. Probing how these features map across the cortex can provide new insights on the neural representation of semantic knowledge.

We have also shown that the high-quality predictions of neural responses afforded by deep neural networks allow accurate decoding of complex stimuli from observed responses. The resulting decoding performance significantly improves on the performance which can be obtained with other established approaches that do not incorporate mid- to high-level stimulus features \cite{Kay2008, Guclu2014}.

Our use of deep neural networks to probe cortical representations is in line with the emerging use of sophisticated techniques that are rooted in statistical machine learning. For instance, in previous work, we have shown that deep belief networks, which can learn stimulus features in a fully unsupervised manner, allow decoding of stimuli from observed neural responses \cite{VanGerven2010}. Recently, it was shown that performance-optimized hierarchical models can predict single-neuron responses in area IT of the macaque monkey \cite{Yamins2013}. Our current work significantly expands on these important results in (i) showing that there is an explicit gradient for object complexity in the ventral pathway of the human brain, (ii) providing an explicit visualization of features in deep layers of a neural network that are subsequently mapped across cortex and (iii) demonstrating that deep neural networks allow decoding of representations in the human brain at a previously unattainable degree of accuracy, providing a sensitive window into the human brain.

\bibliographystyle{ieeetr}
\bibliography{bibliography.bib}
\end{document}